\documentclass
{aa}
\usepackage{txfonts}
\usepackage{fleqn}

\usepackage{url}
 
\usepackage{graphicx}
 
\usepackage{natbib}
\bibpunct{(}{)}{;}{a}{}{,}
\begin{document}


\title{Gravitational lensing as a probe of compact object population in the Galaxy}

\titlerunning{Gravitational Lensing ... }

\author{S. Os{\l}owski\inst{1}
       R. Moderski \inst{2}
        T.  Bulik\inst{1,2}   \and
        K. Belczynski\inst{3,4} 
 }

\offprints{T. Bulik, \email{tb@astrouw.edu.pl}}
\authorrunning{S. Os{\l}owski et al. }

\institute{Astronomical Observatory, 
Warsaw University, Aleje Ujazdowskie 4, 00478  Warsaw, Poland 
            \and 
Nicolaus Copernicus Astronomical Center, Bartycka 18, 00716 Warsaw, Poland 
            \and Department of Astronomy, New Mexico State University, Frenger Mall, Las Cruces, NM USA
	 \and{Tombaugh Fellow}
             }

\date{Received date / Accepted date }

\abstract{}{The population of solitary compact objects in the Galaxy
is very diffcult to investigate. In this paper we analyze the possibility
of using microlensing searches to detect and to analyze the properties of 
the solitary black holes and neutron stars.}
{Evolution of single and binary stars is considered using 
the StarTrack 
population synthesis code. We investigate the properties of the Galactic population 
of compact objects numerically.}
{We find that the compact object lensing events are concentrated in a region 
with the radius of $\approx 5$ degrees around 
the Galactic center. The distribution of masses of the lenses 
for the models we consider  differs but only slightly from the underlying mass
distribution. The expected detection rates are of the order of a few per year.}{}

\maketitle

\section{Introduction}

The stellar evolution models predict the Galactic population
of compact objects - neutron stars and black holes
- numbers somewhere between $10^7$ and $10^9$ objects.
Most of them are solitary  and only a small fraction 
resides in binaries. A small fraction of solitary neutron stars is visible as
radio pulsars.
Some  black holes in 
binaries are actively accreting and are 
observable mainly as X-ray transients, however most of the black hole population has not been yet seen.
The properties of 
black holes population in our galaxy depend on the
history of star formation rate, evolution of metallicity and on the initial 
mass function and on details of compact object formation in supernovae explosions.

The properties of the population of solitary black holes 
can only be investigated indirectly, through observations of their 
interaction with the interstellar matter or light emitted by stars.
Solitary black holes should be accreting gas from the ISM. 
Therefore some of them should be observable
 in  X-rays \citep{2002MNRAS.334..553A,2005A&A...440..223B}.
The luminosity in X-rays shall on one hand be smaller than for solitary 
neutron stars where the surface emission plays a significant role, but on the other hand it may be increased due to smaller velocities and higher
mass of black holes in comparison with neutron stars. The searches for such 
objects have not been successful yet.

Solitary black holes are also detectable with current microlensing searches\citep{Pacz2003}
like OGLE \citep{2003AcA....53..291U} and MOA \citep{2001MNRAS.327..868B}.
These campaigns have already yielded several candidate
black holes detections. \citet{2002ApJ...579..639B} has 
presented two events: MACHO-96-BLG-5 and MACHO-98-BLG-6, with the mass
estimates $6^{+10}_{-3} M_\odot$ and $6^{+7}_{-3} M_\odot$ respectively.
\citet{2002MNRAS.329..349M} showed that 
OGLE-1999-BUL-32 identified also as MACHO-96-BLG-22
is a black hole candidate with the minimum mass of $10.5 M_\odot$.
A further search for X-rays from 
MACHO-96-BLG-5 \citep{2005ApJ...631L..65M} yielded an upper 
limit corresponding to the luminosity of less than $(8-9)\times 10^{-10}\,L_{Edd}$. A recent likelihood analysis of 22 microlensing events 
\cite{2005ApJ...633..914P} lead to  confirmation  of the 
black hole candidate MACHO-99-BLG-22, while the other candidates
are less probable.

The sample of the microlensing compact objects and black hole candidates 
will increase with time. It is therefore interesting to 
see what constraints can be imposed on the models of 
black hole formation and evolution by these observations.
In this paper we present  a simulation   of the 
stellar evolution leading to production of solitary black holes 
in our galaxy. We analyze two scenarios: the single stellar evolution and the
fromation of solitary black holes through disruption of 
binaries. We then analyze the motion of  black holes in the
galactic potential and search for possible microlensing observable from the Earth. 

In section 2 we present the basic ingredients of the calculation:
the stellar evolution model, the galactic potential and mass distribution used, and the lensing search algorithm. Section 3 contains the results and 
section 4 the discussion.

\section{Description of the model}

\subsection{Compact object formation}

The single star evolution is modeled using the 
formulae of \citet{2000MNRAS.315..543H} and we use the {\tt StarTrack}
population synthesis code \citep{2002ApJ...571..394B} for modelling binary evolution.  The {\tt StarTrack}
population synthesis code was initially developed for the study of
double compact object mergers in the context of GRB progenitors  \citep{2002ApJ...571..394B}
 and gravitational-wave inspiral sources
\citep{2002ApJ...572..407B}. In recent years {\tt StarTrack} has
undergone major updates and revisions in the physical treatment of
various binary evolution phases. The new version has already been
tested against observations and detailed evolutionary calculations
\citep{Belcz2006}, and has been used in various applications.
 The most important updates for compact
object formation and evolution include: a full numerical approach to
binary evolution due to tidal interactions and coupling calibrated
using high mass X-ray binaries and open cluster observations, a
detailed treatment of mass transfer episodes fully calibrated against
detailed calculations with a stellar evolution code, updated stellar
winds for massive stars, and the latest determination of natal kick
velocity distribution for neutron stars \citep{2005MNRAS.360..974H}. 
For disrupted (by supernova explosion) binaries we follow trajectories of 
components as described in \citet{2006ApJ...650..303B}.
In the helium star evolution, which is of a crucial importance for the
formation of new classes of double compact objects, e.g., \citet{2003ApJ...592..475I}, we have applied a conservative treatment matching closely
the results of detailed evolutionary calculations. The NS-NS
progenitors are followed and checked for any potential Roche lobe
overflow (RLOF). While in the mass transfer phase, systems are examined for
potential development of dynamical instability, in which case the systems
are evolved through a common envelope phase.
We treat common envelope events through the energy formalism 
\citep{1984ApJ...277..355W,2002ApJ...572..407B}, where the binding energy of the envelope
is determined from the set of He star models calculated with the
detailed evolutionary code by \citet{2003ApJ...592..475I}.
For some systems we observe, as before, extra orbital decay leading to
the formation of very tight short lived double compact object binaries.
However, since the progenitor evolution and the final RLOF episodes
are now followed in much greater detail, we note significant
differences from our earlier studies. For a detailed description of the
revised code we refer the reader to \citet{Belcz2006}.

\begin{table}
\caption{Models of black hole formation considered in this paper}
\begin{tabular}{ll}
Model   & Description \\ \hline \hline
A       & Standard\\
W05     & stellar winds decreased by 0.5\\
K0      & Black holes receive no kicks \\
K1      & Black holes receive full kicks \\
C       & Hertzsprung gap stars can be donors in CE phase\\
S       & Only single stars \\
M       & Black holes from mergers \\
\end{tabular}

\label{models}
\end{table}

The simplest scenario leading 
to a single compact object is through evolution of single stars.
In this case we put the newly formed star in the galactic
disc with its velocity drawn from an appropriate distribution
after the supernova event.

An important additional scenario follows a disruption 
of a binary  as a result of a first supernova explosion. 
A large fraction of binaries containing massive stars 
is disrupted in this way.  This leads to a formation of a single 
compact object and a single companion which may still be massive enough
to produce another compact object. We assume that the first 
supernova explosion takes place in the galactic disk. We follow 
the motion of the single star   until it 
produces another compact object.
 
A much smaller contribution 
comes from the disruption of a binary  during the second 
supernova explosion. In this case two single compact objects are
formed.  We follow the motion of the system after 
the first supernova until the second one. Thus the newly formed 
single compact objects start at the actual location of the second supernova
with appropriate velocities.

Finally, we also investigate black hole formation through mergers. This includes mergers 
of stars during their nuclear evolution, as well as mergers of
a compact object with a massive companion as a result 
of binary interaction. In the first case we assume that a single star is formed
with the mass equal to the sum of the masses of the two merging
components and we evolve this star neglecting the potential changes in metallicity. In the latter case we assume that a single black hole is formed with the mass equal to the sum of the masses of the compact object and the core of the companion while the envelope is expelled. The newly formed black holes do not gain additional velocities 
during mergers. 

We ignore the contribution of double black hole systems, and 
single black holes formed in mergers of double black holes as they
represent a much smaller population than the ones mentioned above.

We consider several models of binary evolution to asses the 
sensitivity of the results. We decrease the  strength of the stellar winds 
which  affects mainly the masses of the newly formed compact objects (model W05).
A second parameter that may influence the results significantly 
is the distribution of the kicks received by the newly formed compact object.
Apart from the standard model in which the value of the kick is decreased 
with increasing fall-back mass, we consider two extreme cases: in first black holes 
receive no kicks at all (model K0), and in the second one kick distribution regardless of their mass is the 
same as for neutron stars (model K1).
Finally in model C we allow for the common envelope evolution initiated 
by stars passing through the  Hertzsprung gap.

The list of the models 
considered is shown in Table~\ref{models}.

\subsection{Galactic model}

We consider a model of the Galaxy consisting of three
components:   bulge,   disk, and  halo. 
The bulge and disk potential are described by 
the \citet{1975PASJ...27..533M} type potential
\citep{1990ApJ...348..485P,1999MNRAS.309..629B}
\begin{equation}
\Phi(r,z)= \frac{GM}{\sqrt{R^2+(a+\sqrt{z^2+b^2})^2} } \, ,
\end{equation}
where $M$ is the mass of a given component, $R=\sqrt{x^2+y^2}$,
and $a,B$ are the parameters.
The halo is described by the density distribution 
$\rho=\rho_c [1+(r/r_c)^2]^{-1}$ with a cutoff at $r_{cut}=100$\,kpc above which 
the halo density is zero. The corresponding potential for $r<r_{cut}$
is
\begin{equation}
\Phi(r)= - \frac{GM_h}{r_c} \left[ \frac12 \ln\left(1+ \frac{r^2}{r_c^2} \right) + \frac{r_c}{r} \arctan\left(\frac{r}{r_c} \right) \right]
\end{equation}
We use the following values of the parameters \citep{1991ApJ...381..210B}
describing the potential: $a_1=0$\,kpc, $b_1=0.277$\,kpc, $a_2=4.2$\,kpc, $b_2=0.198$\,kpc, $M_1=1.12\times 10^{10} \,M_\odot$, 
$M_2=8.78\times 10^{10} \,M_\odot$, $M_h=5.0\times 10^{10} \,M_\odot$, and
$r_c=6.0$\,kpc.

 The distribution of stars in the disk is assumed to be 
that of a young disc \citep{1990ApJ...348..485P}. The 
radial and vertical distributions are independent i.e.
the distributions factor out:
\begin{equation}
P(R,z)\propto R(R)dR \, p(z) dz \, ,
\end{equation}
where the radial distribution is
$p(R) \propto \exp(-R/R_{exp})$, and $R_{exp}= 4.5\,$kpc, and we introduce
an upper cutoff at $R_{max}=20$\,kpc. The vertical distribution is exponential 
$p(z) \propto \exp\left(-z/75{\rm pc}\right)$ .
This is not a self consistent approach since the matter density corresponding  to the disk potential is not the same as the 
stellar density. A self consistent approach has been 
investigated by \citet{1998ApJ...505..666B} using the potentials calculated 
by \cite{1989MNRAS.239..571K}.

\subsection{Lens search algorithm}
A naive search for lenses i.e.\ testing for events when a star and a
compact object are aligned with Earth on its galactic orbit would
require nearly infinite computational resources.  Therefore we use the
following lens search algorithm. We look for the lens events that
take place anywhere on the galactic orbit of the Earth.  This is much
easier computationally as this approach make use of the cylindrical
symmetry of the Milky Way.
\begin{figure} 
  \begin{center}
  \includegraphics[width=\columnwidth]{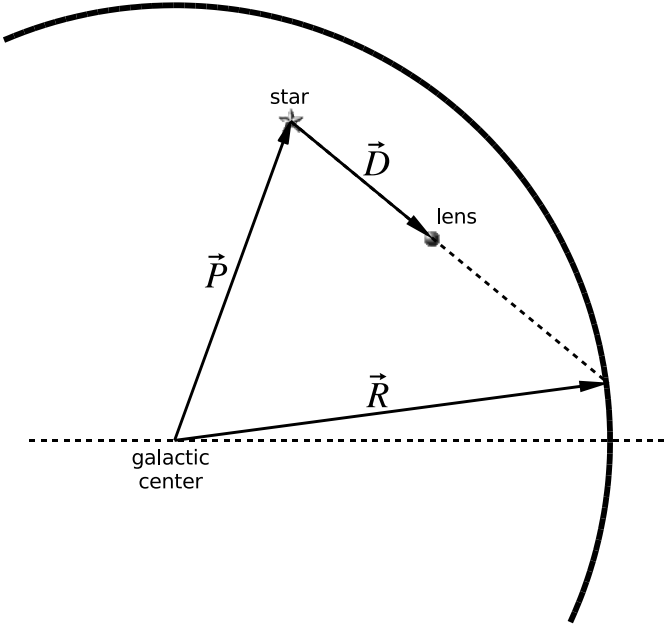}
  \end{center}
\caption{Geometry of the lensing event. Star and the lens are located
  inside a sphere of radius $R$ equal to the radius of the Earth's
  orbit.  Vector $\vec{P}$ denotes the position of the star in the
  galaxy, while $\vec{D}$ is a star-lens vector. We look for a
  situation when star-lens line crosses the Earth's orbit during the
  evolution of the system.}
\label{fig:lens}
\end{figure}

We assume that the Earth's orbit is circular with the radius of
$8.5$\,kpc.  Let us consider a sphere of radius $R=8.5$\,kpc (which is
equal to Earth's galactic orbit radius) and centered in the center of
the galaxy.  Let us also consider a star-lens system located somewere
in the galaxy.  The position of the system is characterized by two
vectors: $\vec{P}$ - the position vector of the star in the galaxy,
and $\vec{D}$ - the star-lens vector (see Fig.~\ref{fig:lens}).  We
search for a point $Q$ in which the line passing through the position of
the star and the lens pierces the sphere.  In principle we look for
a factor $t$ which satisfies the condition
\begin{equation}
\left \vert \vec{P} + t\vec{D} \right \vert = R \ .
\label{eq:ptdr}
\end{equation}
Eq.~(\ref{eq:ptdr}) is a quadratic equation in $t$
\begin{equation}
\left( \vec{P} + t\vec{D} \right) \cdot \left( \vec{P} + t\vec{D}
\right) = R^2 \ ,
\end{equation}
\begin{equation}
D^2 t^2 + 2\vec{D} \cdot \vec{P} t + (P^2 - R^2) = 0 \ .
\label{eq:t}
\end{equation}
Providing that $\Delta = 4 \left( \vec{D} \cdot \vec{P} \right)^2 - 4
D^2 \left(P^2 - R^2 \right) \geq 0$ the Eq.~(\ref{eq:t}) has a
formal solution:
\begin{equation}
t_{1,2} = \frac{1}{D^2} \left[ -\vec{D} \cdot \vec{P} \pm \sqrt{
\left( \vec{D} \cdot \vec{P} \right)^2 - D^2 \left( P^2 - R^2 \right)}
\right] \ .
\label{eq:t12}
\end{equation}

Now, for each pair of a star and a compact object, we calculate
$\Delta$ taking into account their positions at the beginning and at the end of a
calculation step.  If $\Delta>0$ we calculate $\Delta'$ from the
position of the pair at the end of the step, which in our calculations was one
month.  If it is also positive, then we calculate both $t_{1,2}$ and
$t'_{1,2}$, respectively.  We are only interested in a configuration
for which $t_{1,2}>1$ and $t'_{1,2}>1$.  If $t<0$ the compact object
is located behind the star,  and for $0 \le t \le 1$ the compact object and the 
star reside on the opposite sides of the sphere.  In both such cases the lensing event
cannot be observed on Earth.  For both $t$ and $t'$ we then calculate
positions of the points where star-lens line crosses the sphere at the
beginning and at the end of the examined evolution period.  The lensing
event happens if these points are located in different hemispheres ($z'z<0$).

This algorithm is not sensitive to some special configurations when
during the evolution star or compact object are located outside the
sphere at very high azimuthal angles (as seen from Earth).  We checked
for these special occasions with two additional algorithms.
We conclude that such events are extremely rare (there was not even one such case) and does not influence
our results.

\subsection{Lensing rate}

In order to estimate the true lensing rate one has to scale 
the number of the lensing events found in the simulation  to match the physical conditions in our Galaxy.  There are
two scaling factors involved. One results from the number of simulated
stars and compact objects and the second from the fact that we look
for the lensing events that happen on the whole Earth's orbit.
Let us denote the Earth orbital period in the Milky way as 
$P=250$\,Myrs.Let 
the true number of star in the Milky Way be $N^*_{11}\times 10^{11}$,
and $f_{lens} =10^{-2} f_{lens,-2}$ represent the fraction of the stars in the galaxy that 
can be detected by lens search experiments.  
and the number of compact objects in the Galaxy $N^{CO}_8\times10^8$.
Our simulations include  $10^6$ stars that can be lensed in the Galactic potential.

Thus the expected rate of lenses due to the compact 
objects is 
\begin{equation} 
R  =  4\times 10^2 f_{lens,-2} N^*_{11} N^{CO}_8 \frac{N_{lens}}{N_{sim}}\, ,
\label{rate}
\end{equation}
where $N_sim$ is the number of compact objects (potential lenses)
in the simulation, and $N_{lens}$ is the number of lenses we
find in one year.  

\begin{figure}
\includegraphics[width=\columnwidth]{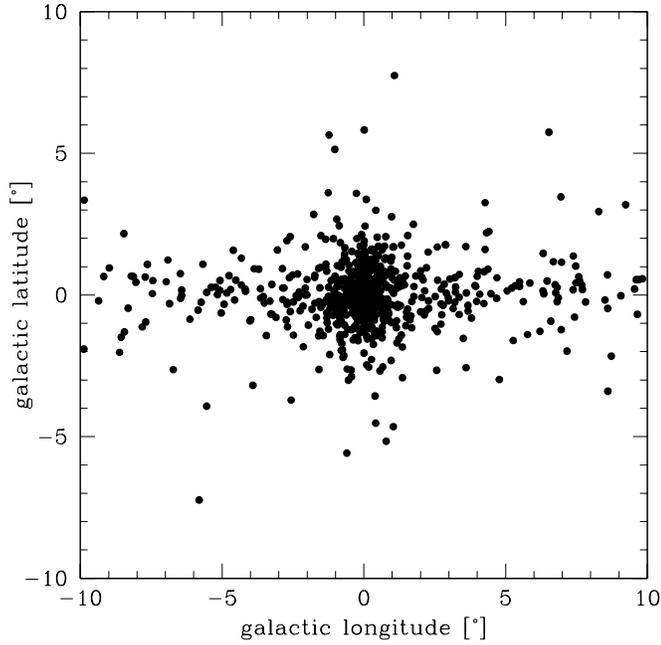}
\caption{Skymap  of the sumlated lensing events  for the
standard model of compact object formation. The skymaps 
in the two alternative models look similar.}
\label{maps}
\end{figure}

\section{Results}

\begin{table}
\caption{The results of the lens search simulations and the expected 
galactic lensing rates.}
\begin{tabular}{llll}
Model &    & \\
       & $N_{sim}$ & $N_{lens}$ & Rate[yr$^{-1}$] \\ \\ \hline\hline 
A      & 76399          & 826   & 4.3                \\
W0.5    & 73575         & 757   & 4.1              \\
K0	& 26619		& 584	& 8.7 \\
K1	& 73907		& 687	& 3.7 	\\
C	& 81327		& 887	& 4.3 	\\
S	& 60340		& 1368	& 9.0 	\\
M	& 8170		& 194	& 9.5	\\
\end{tabular}
\end{table}

Using Startrack population synthesis code we evolved $10^5$ binary systems, obtaining data about compact objects (neutron stars and 
black holes). We put those objects (only solitary, from disrupted systems)
 into the model Galaxy.We also 
randomly introduced $10^6$   stars in the Milky Way. All the stars
 were then evolved \ for $10Gy$ in Galactic potential taking into account 
the initial  velocities that the compact objects received  
at birth due to asymmetric kicks and disruptions of the binary 
systems \cite{1961BAN....15..265B}. We then searched for 
lensing events that take place anywhere on the Earth orbit in the 
galaxy in one year of simulation.  
The resulting rates, calculated using equation \ref{rate}, 
 are presented in Table~1 for each model 
assuming that $N^*_{11}=1$, and $N^{CO}_8=1$. 
The typical values obtained are a few tens per year, however
these expected rates have to be considered as rough 
estimates given the number of assumptions used in equation~\ref{rate}.
Note that the Table gives the values of the rates assuming that 
all compact objects are formed in the particular scenario.

\begin{figure}
\includegraphics[width=\columnwidth]{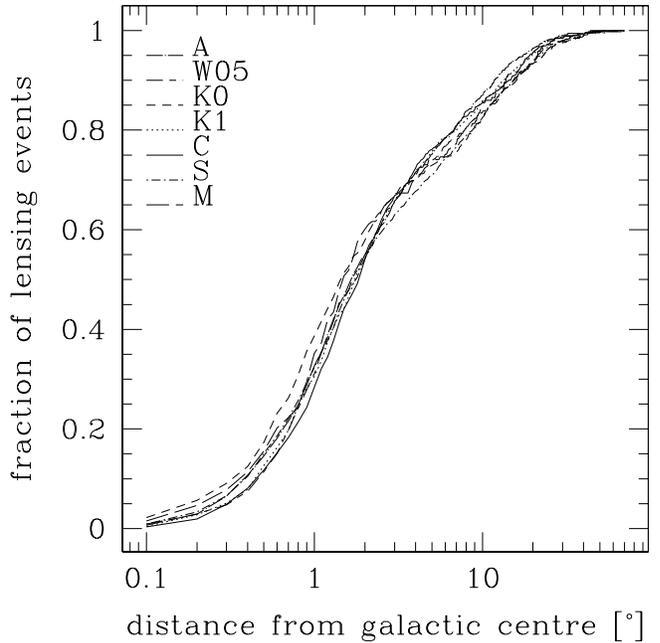}
\caption{Cumulative distributions of the angular distance
 lensing events from the Galactic Center for the models considered in the paper.}
\label{circles}
\end{figure}

\begin{figure}

\includegraphics[width=0.9\columnwidth]{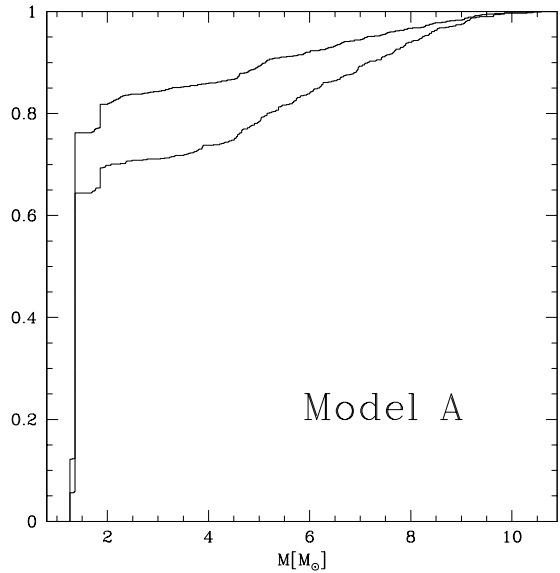}

\caption{The cumulative 
distributions of masses of compact objects
in the standard model A.
The thick line denotes 
the intrinsic distribution of compact object masses
while the thin line corresponds to the distribution 
of measured masses of the lenses. 
each plot is labeled by the model as in Table 1.}
\label{masses}
\end{figure}

In each simulation we have noted the positions of the lensing events on the sky. We present the skymaps of these position in Figure~\ref{maps}. The 
lenses are strongly concentrated around the galactic center. In Figure~\ref{circles} we present the cumulative distribution of the 
fraction of events as a function of the distance from the Galactic 
Center. Typically $70$\% of the lensing events take place within 
$\approx5^\circ$ from the Galactic 
Center, while  a fraction $0.9$ of the events happens within $20^\circ$.

\begin{figure*}
\includegraphics[width=0.3\textwidth]{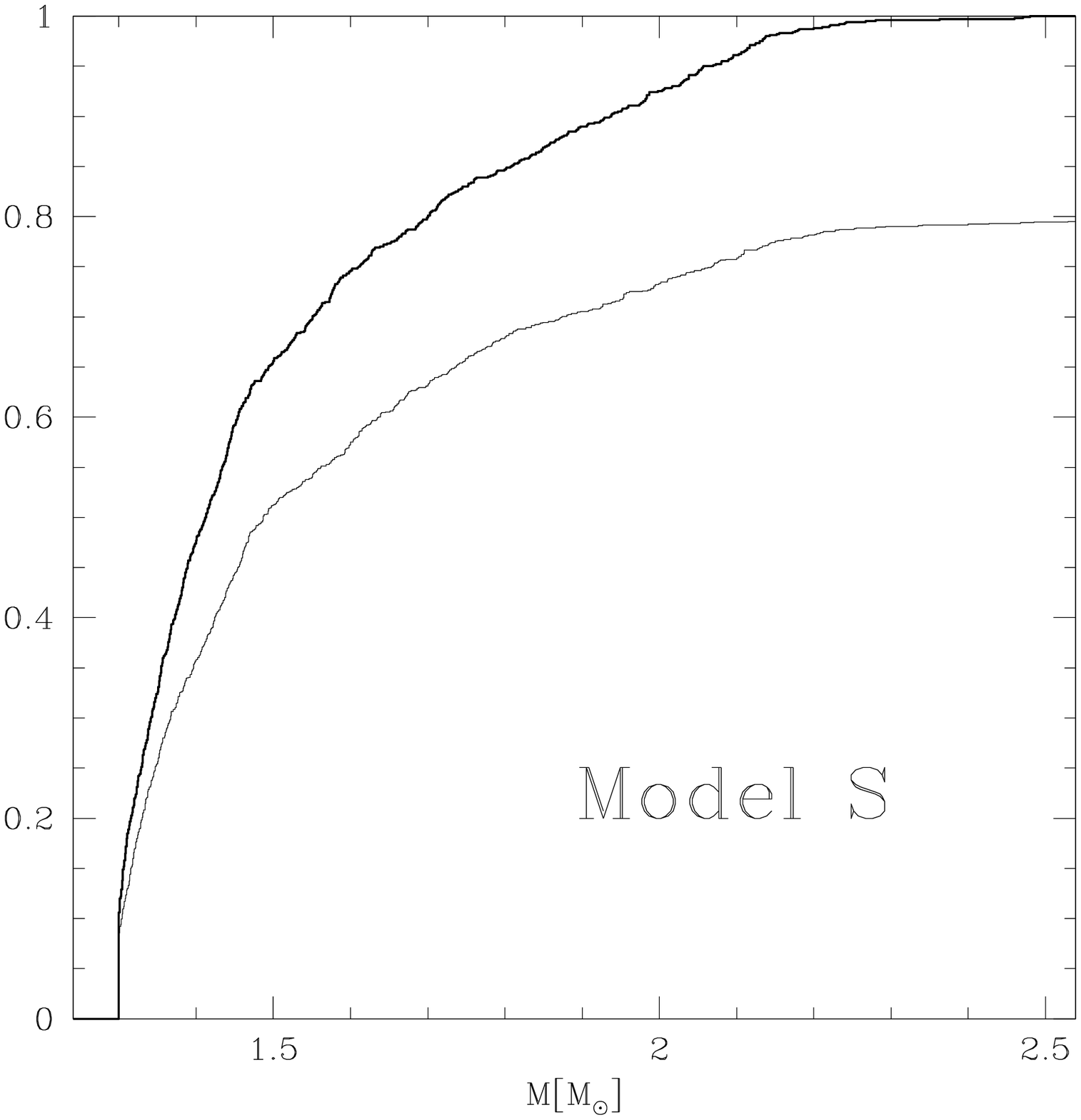}
\includegraphics[width=0.3\textwidth]{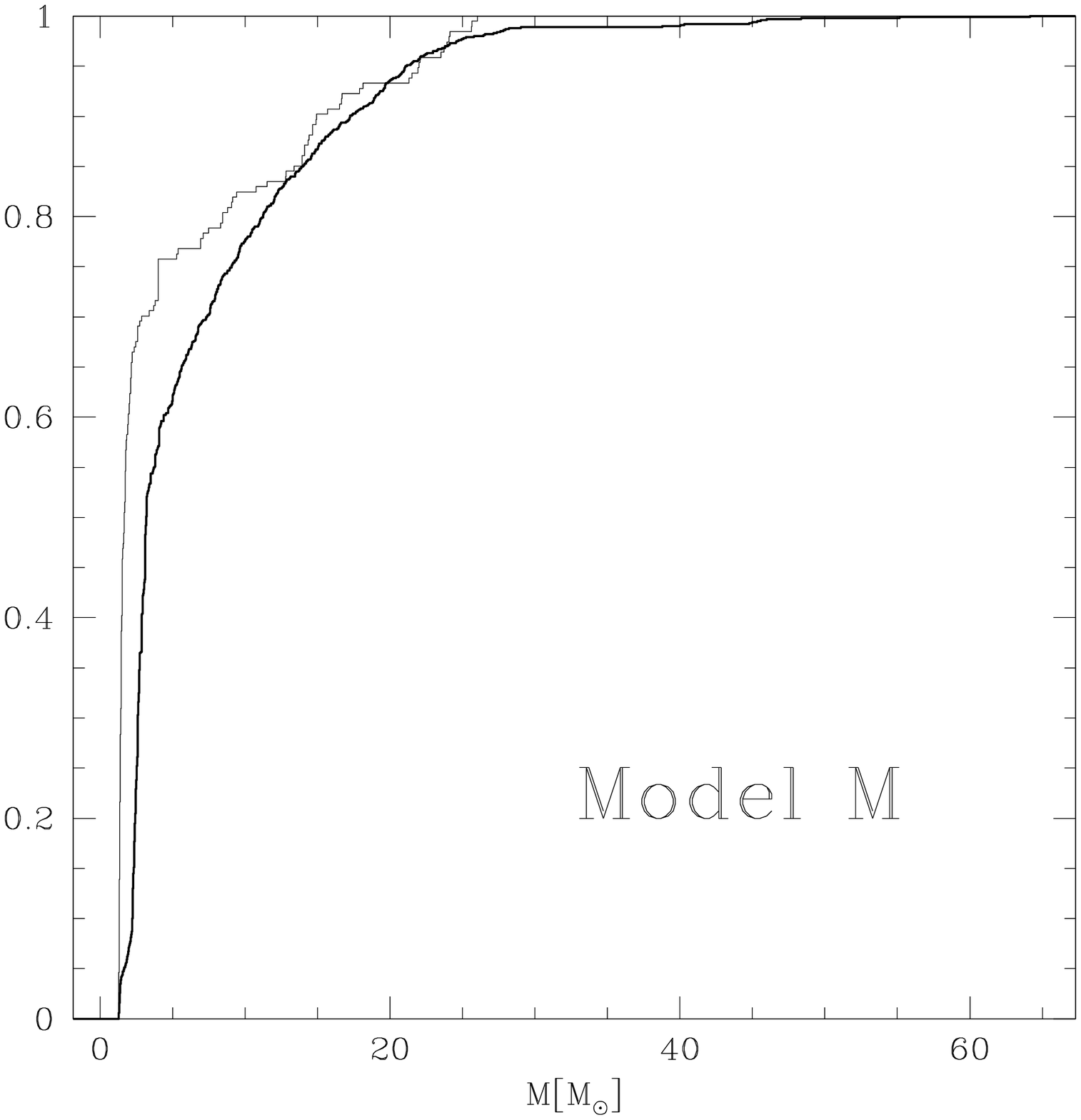}
\includegraphics[width=0.3\textwidth]{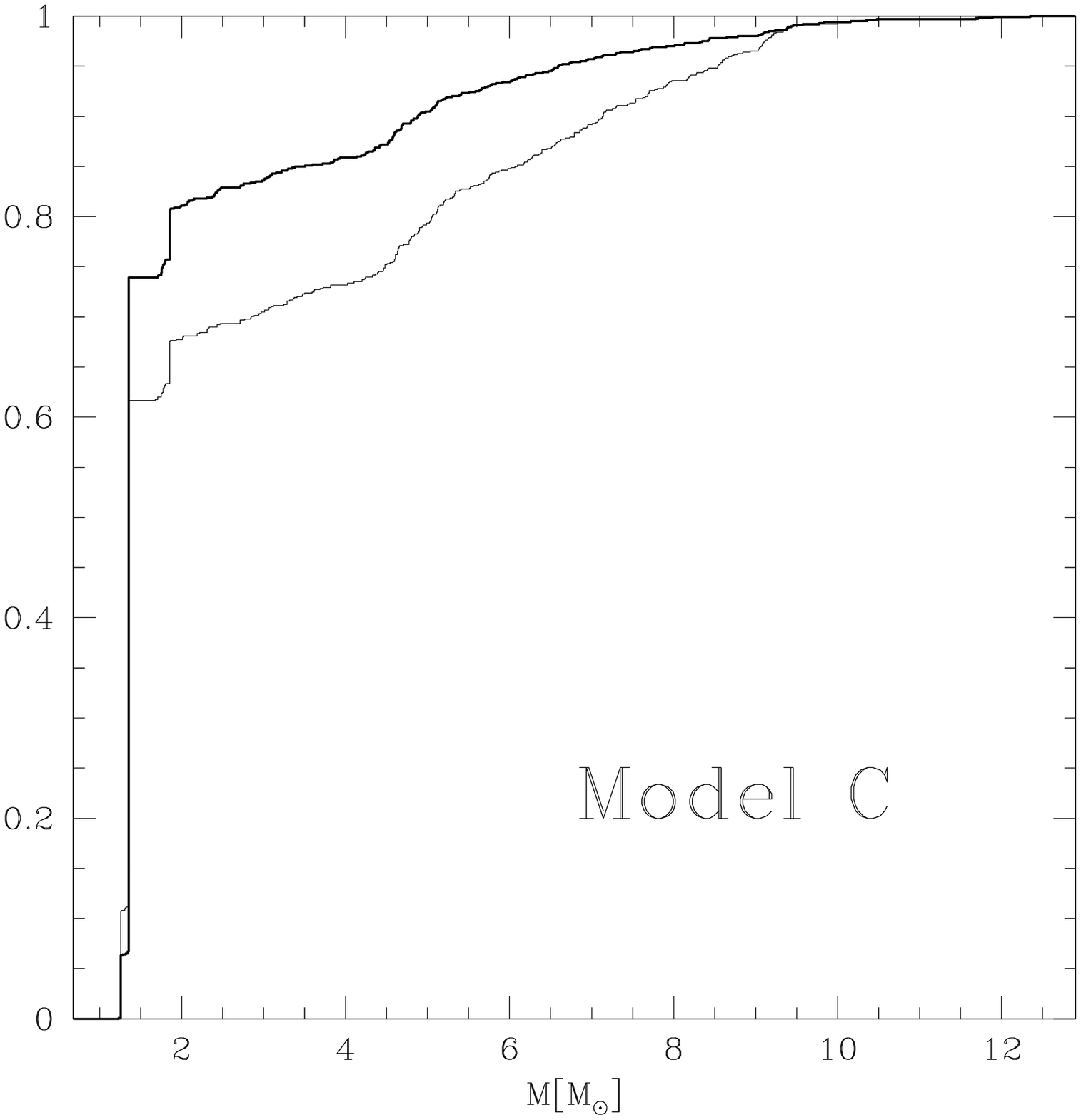}\\
\includegraphics[width=0.3\textwidth]{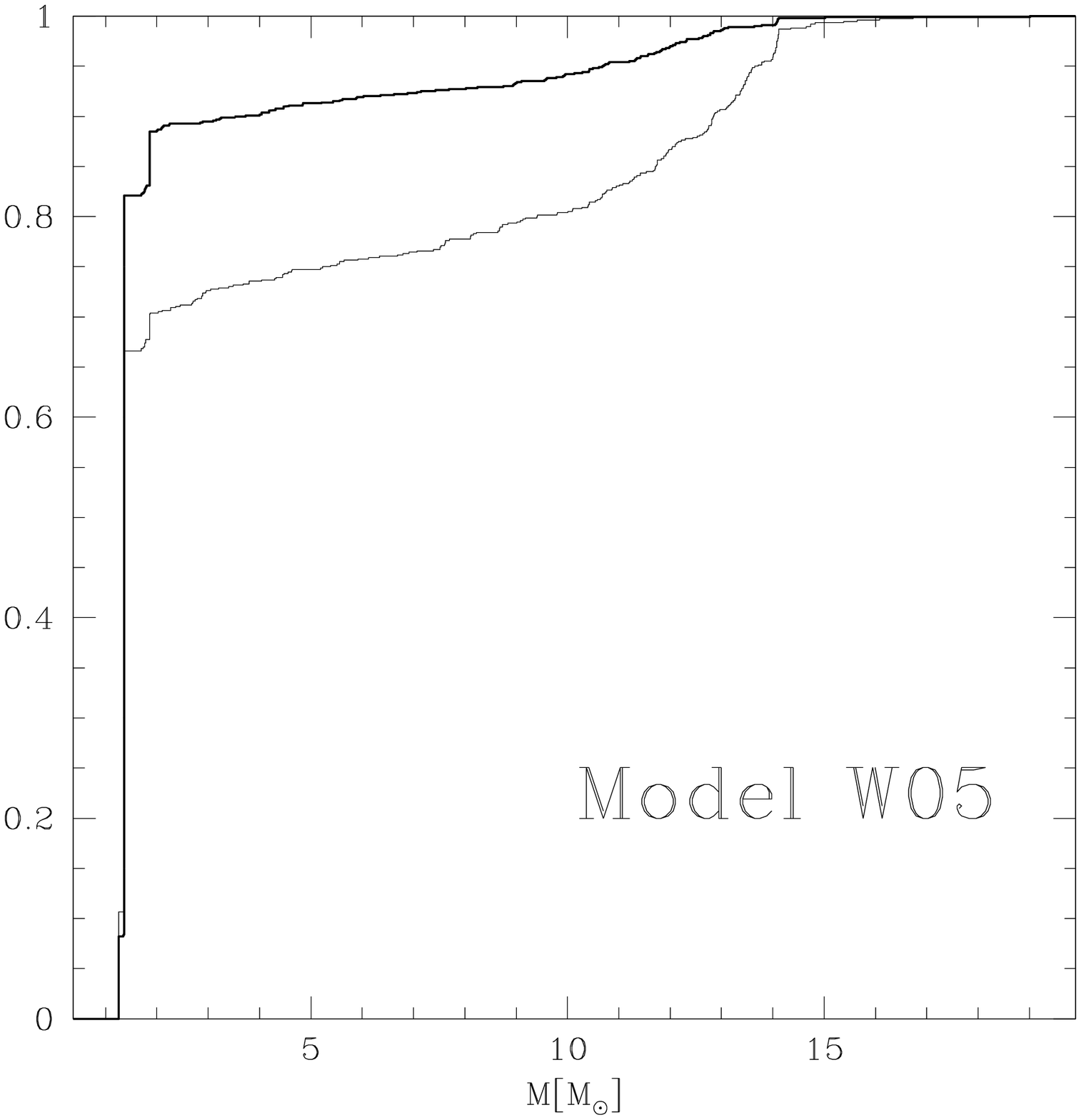}
\includegraphics[width=0.3\textwidth]{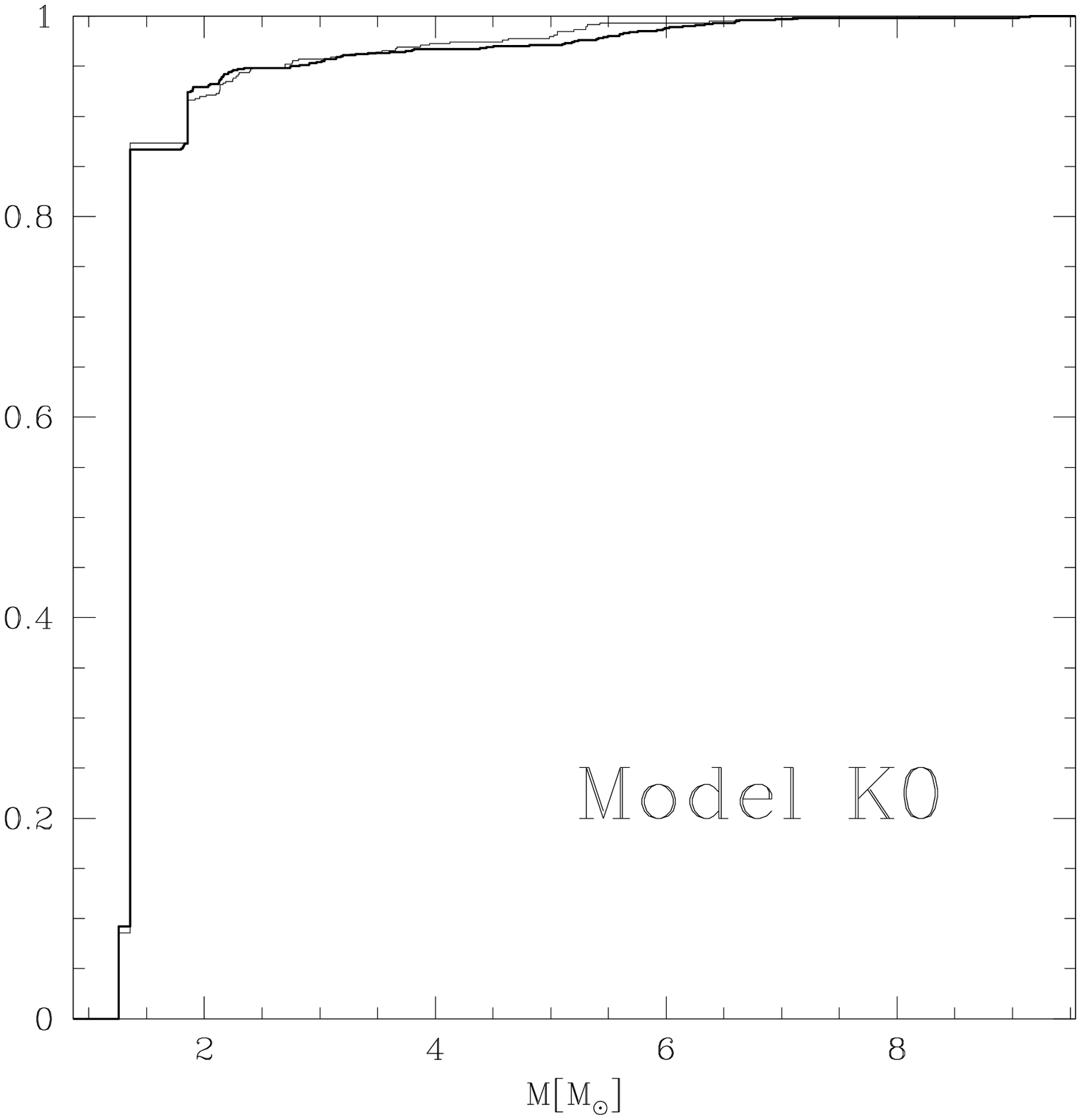}
\includegraphics[width=0.3\textwidth]{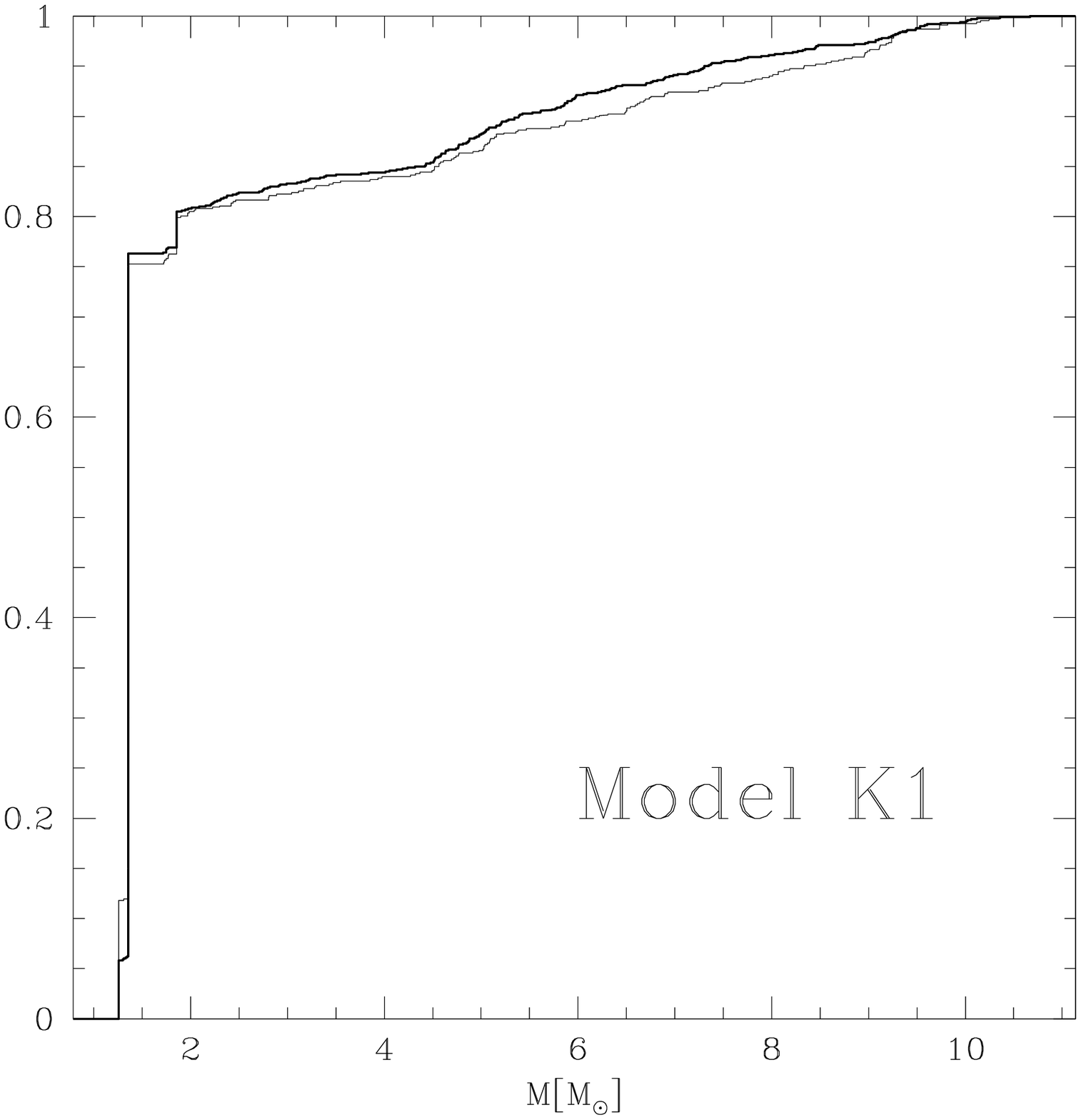}\\

\caption{Same as Figure 4 but for the remaining models. }
\label{massestwo}
\end{figure*}

We also note the mass of the gravitational lens 
for each lensing event. In Figures~\ref{masses} and \ref{massestwo}
 we present 
the  intrinsic distributions of masses of compact objects 
for each model and the mass distribution of lenses.
In the case of models S, M, K0, and K1 the distributions are either quite 
similar. They do differ in the case of remaining models. 
But if they differ then they do in the same manner:the distribution of observed masses shows a deficiency 
of low mass compact objects (neutron stars) and an increased fraction 
of high mass objects (black holes).  The distributions are similar for the models
in which the velocities of compact objects weakly depend on their masses:
K0,  and K1.

In the cases when the distributions differ the suppresion of 
low mass compact objects is not more than about 20\% in relation to the high 
mass compact objects.

\section{Summary}

We have considered several scenarios leading to formation of galactic
compact objects and calculated their initial velocity distributions.
We then simulate the spatial and velocity distributions of 
them in our Galaxy. Each model yields 
a different compact objects mass distribution  
and may have a different distribution of velocities. 
We simulate the microlensing events and note 
both mass and position for each one.

We estimate the rates of lensing events due to 
compact objects. These estimates show that they can 
be observed. The rates in Table 2 must be taken
as very rough estimates. There are some poorly known 
factors that are included in these rates - see equation~\ref{rate}.
One should note that the fraction of objects detectable by the 
lens searches $f_{lens}$ can be increased for the searches conducted in the
infrared.
However, we know that there already are some events which 
can be interpreted as microlensing events by black holes.

In the simulations the robust result is the distribution
of lensing events in sky. All the events are well
concentrated around the Galactic center. Thus the lens searches
should concentrate on this region. This is probably due to the fact 
that the dominant factor determining the position on the sky 
is the density of the lensed stars and not lenses themselves.

Finally, we find that the observed mass distributions
differ by less than 20\% from the intrinsic one. 
The observations of several gravitational lenses 
from black holes or neutron stars should lead to 
relatively good estimate of the intrinsic 
compact object mass distribution. The observation of 
the mass distribution of compact objects is extremely
important since this is a unique way to probe the 
final stages of the massive stars evolution. Observations
of black holes in massive accreting binaries may suffer 
from numerous selection effects and probably do not 
reflect the underlying intrinsic distribution.

The lensing events due to massive lenses will
have long duration and therefore require 
long observational campaigns. We emphasize 
the importance of such searches, as their
potential results will have a very significant 
relevance for the theory of compact object formation
and models of massive star evolution.

\begin{acknowledgements}
This research  was supported by the KBN grant 1P03D02228.Te authors thank
A. Udalski for useful commets on the manuscript.
We are also grateful to A. Sadowski for help with the  merger calculation.
\end{acknowledgements}


\end{document}